\documentclass[11pt]{article}
\usepackage[top=0.7in, bottom=0.7in, left=0.7in, right=0.9in]{geometry}
\usepackage{float}
\usepackage{pgf, tikz}
\usetikzlibrary{arrows, automata}

\usepackage[textwidth=8em,textsize=small]{todonotes}
\usepackage[utf8]{inputenc}
\usepackage{mathbbol}
\usepackage{bm}
\usepackage{url}
\usepackage{amsmath}
\usepackage{natbib}

\usepackage{booktabs} 
\usepackage{changepage}
\usepackage{color}
\usepackage{graphicx}

\begin{document}
\begin{center}
{\Large Statistical implications of relaxing the homogeneous mixing assumption in time series Susceptible-Infectious-Removed models}\\ \ \\

Luis D. J. Martinez Lomeli$^1$, Michelle N. Ngo$^1$,
Jon Wakefield$^2$,
Babak Shahbaba$^{1,3,*}$, Vladimir N. Minin$^{1,3,*}$ \\[4pt]
$^1$Center for Complex Biological Systems, University of California, Irvine. \\
$^2$Departments of Biostatistics and Statistics, University of Washington, Seattle, Washington\\
$^3$Department of Statistics, University of California, Irvine\\[2pt]
$^*$corresponding authors:  \url{babaks@uci.edu, vminin@uci.edu}
\end{center}



\begin{abstract}{
Infectious disease epidemiologists routinely fit stochastic epidemic models to time series data to elucidate infectious disease dynamics, evaluate interventions, and forecast epidemic trajectories. 
To improve computational tractability, many approximate stochastic models have been proposed.
In this paper, we focus on one class of such approximations --- time series Susceptible-Infectious-Removed (TSIR) models.
Infectious disease modeling often starts with a homogeneous mixing assumption, postulating that the rate of disease transmission is proportional to a product of the numbers of susceptible and infectious individuals. 
One popular way to relax this assumption proceeds by raising the number of susceptible and/or infectious individuals to some positive powers. 
We show that when this technique is used within the TSIR models they cannot be interpreted as approximate SIR models, which has important implications for TSIR-based statistical inference. 
Our simulation study shows that TSIR-based estimates of infection and mixing rates are systematically biased in the presence of non-homogeneous mixing, suggesting that caution is needed  when interpreting TSIR model parameter estimates when this class of models relaxes the homogeneous mixing assumption. 
}
\end{abstract}


\section{Introduction}
\indent \par Susceptible-Infectious-Recovered (SIR) models, first presented by \cite{Kermack1927}, are commonly used to capture the dynamics of an infectious disease outbreak in a population.
Stochastic versions of SIR-like models tend to be difficult to fit due to computational intractability of the likelihood function.
One solution to circumvent this issue is to use time series SIR models (TSIR) \citep{finkenstadt2000dynamical,becker2017tsir, baker2019, Giles2020, giles2021}, which are based on a branching process approximation of the epidemic progression.
Although the branching process approximation of the TSIR model has a rigorous mathematical justification, its validity breaks down when the assumption of homogeneous mixing (also known as the law of mass action) is relaxed. 
This could lead to an unjustifiable model misspecification. 
In this paper, we investigate the statistical implications of such model misspecification and find that the TSIR model with non-homogeneous mixing can produce biased estimates of the infection rates.

Fitting stochastic epidemic models to empirical data within a likelihood-based framework is a challenging task because it requires integrating over a large number of missing data, which is computationally burdensome.
Methods that approximate the likelihood function improve the computational tractability have a long history in infectious disease modeling.
Discrete time models that move individuals across multiple compartments provide one class of such approximate methods \citep{longini1982household,lekone2006statistical,held2012modeling}. 
While these approximate models are  tractable, they rely on several simplifying assumptions that make them harder to apply to more general settings, for example, when data are collected at uneven observation times.
Particle filter optimization and Markov chain Monte Carlo (MCMC) approaches offer another solution to the intractable likelihood problem in maximum likelihood and Bayesian framework  \citep{andrieu2010particle,ionides2011iterated, ionides2015inference, dukic2012tracking,koepke2016predictive}.
However, particle filter methods are computationally expensive and can suffer from particle impoverishment problems.
Although more general data augmentation MCMC approaches have been successfully applied to epidemic models, in some cases, it is necessary to impute more unknown variables than is feasible in practice \citep{cauchemez2004bayesian,jewell2009bayesian,o2010introduction,fintzi2017efficient}. 
Finally, problems with intractable likelihoods can be tackled by Approximate Bayesian Computation methods, 
these approaches are computationally intensive, with performance being sensitive to the type of summary statistics used \citep{mckinley2009inference,toni2009approximate,likelihoodfree2015}.
\par
We are interested in the TSIR family of models that provide a simple approximation to the SIR process, allowing for analysis of incidence data (the most common type of surveillance data reported), and remain tractable when extended to simultaneous modeling of multiple geographic locations \citep{xia2004measles, jandarov2014emulating, wakefield2019spatio}.
In particular, the likelihood function takes the form of a product of negative binomial probabilities when the number of infections in a population is approximated by a linear birth process under the assumption of homogeneous mixing \citep{feller1968introduction, allen2010introduction}.
This negative binomial-based approximation is commonly generalized to non-homogeneous mixing, but this generalization has not been rigorously justified and its statistical implications are yet to be explored. 
\par
In this work, we study statistical implications of TSIR under non-homogeneous mixing using 
recent computational advances of fitting Bayesian SIR models to data, to which we refer to as \textit{BayesSIR}  \citep{ho2018direct}. 
The main contributions of the paper are as follows:
(1) We show where the TSIR branching process approximation fails when applied to dynamics with non-homogeneous mixing and provide numerical evidence for the persisting discrepancy between  the transition probabilities of TSIR and SIR models;
(2) we provide a detailed analysis of the model misspecification effect by comparing the performance of BayesSIR to our Bayesian implementation of the TSIR model with the negative binomial-based likelihood, to which we refer as \textit{BayesTSIR}. 
Using synthetic data, we show that disagreement between BayesTSIR-based  and BayesSIR-based inference can be substantial;
(3) we extend the \textit{BayesSIR} model to allow for time-varying infection rates in order to provide a rigorous framework for modeling incidence time series data routinely collected during infectious  disease surveillance. 
We show that the discrepancies between \textit{BayesSIR} and \textit{BayesTSIR} still persist in this setting.

In the last contribution, we apply our generalization of \textit{BayesSIR} with time-varying infection rates to the analysis of the historical London measles dataset that contains biweekly incidence records collected before the implementation of vaccination campaigns.
In this dataset, the measles time series shows a seasonal pattern, where higher case numbers were reported during school-term months compared to school breaks.
To account for this pattern, we consider two ways in which the infection rate can vary throughout the year: (1) with two infection rates, shared across years, corresponding to school-term and school-break, and (2) with rates that vary biweekly throughout each year, but the same biweekly period has the same infection rate across years.
Our results show that infection rate estimates produced by the BayesSIR with non-homogeneous mixing fluctuate less throughout a year than the analogous estimates produced by the BayesTSIR model, suggesting presence of  substantial biases in BayesTSIR-based estimates caused by the TSIR approximation. 



\section{Methods\label{sec:Methods}}

\subsection{Incidence Epidemic Data}
\indent \par Suppose we observe a time series of an infectious disease incidence $\bm{Z}=\{Z_1,...,Z_n\}$, where $Z_i$ represents the number of new cases reported in a period $i$ (e.g., period length could be a week). 
We would like to infer the dynamics of the infectious disease spread. 
For this purpose, we follow the TSIR model assumptions and postulate that all individuals infected in the period $i$ will recover at the end of this period. 
This means that $\bm{Z}$ can be thought of as underreported disease prevalence, defined as the number of infectious/infected individuals in each period.
Therefore, to proceed  with statistical inference  we need a model for changes in the actual prevalence.



\subsection{Stochastic SIR Model}
\indent \par We use a Susceptible-Infectious-Removed (SIR) process to model the prevalence dynamics, including a modification that allows for non-homogeneous mixing. 
More specifically, we define a continuous-time stochastic process $\mathbf{X}(t) =\{S(t),I(t)\}_{t\geq 0}$,  where $S(t)$ and $I(t)$ denote the numbers of susceptible and infectious individuals at time $t$ respectively. 
Given a population of size $N$ with no demographic changes (births, deaths, migration, etc.), and for a small time interval $(t, t +\Delta t)$, the possible events that can occur are (1) a new infection of a susceptible person, (2) a new recovery of an infectious individual, or (3) no change in the population.
These events have the following associated probabilities:
\begin{equation}
\resizebox{0.93\linewidth}{!}{
\def\arraystretch{1.2}
$\mathbb{P}[ \mathbf{X}(t+\Delta t) = (k, l)\; |\; \mathbf{X}(t) = (s, i)  ] =
\left\{
\begin{array}{ll}


 \frac{\beta}{N} \, s \, i^{\alpha} \Delta t + o(\Delta t) &  (k,l)=(s-1,i+1), \\

  \gamma i \, \Delta t + o(\Delta t) &  (k,l)=(s, i - 1), \\

1-(\frac{\beta}{N} \, s\,i^\alpha + \gamma\, i ) \Delta t + o(\Delta t) & (k,l)=(s,i),\\

o(\Delta t) &\textrm{otherwise,}

\end{array}\right.
$}
\label{eq:Stochastic-SIR}
\end{equation}
where $\beta$ and $\gamma$ are the infection and recovery rates respectively, and $\alpha$ represents the mixing parameter for non-homogeneous dynamics.
In the classical SIR model, the mixing rate is assumed to be $\alpha=1$, corresponding to the assumption of homogeneous mixing. 
By setting $\alpha<1$ we make the infection rate slower than it would be in the classical SIR, which is reasonable if we expect that susceptible and infectious individuals are not coming into contact with each other as freely as they would be under the homogeneous mixing assumption. 
Note that the model only keeps track of the number of susceptible and infectious individuals but it is possible to obtain the number of individuals in the recovered compartment, since at all times it is assumed that the three compartment sizes add up to the population size $N$.

We extend the SIR process to model the infection process across multiple periods of time where the infection rate can vary per period.
For example, we can consider a single infection rate per month, season, or year. 
This implies that the vector of SIR parameters across all the observation periods of interest can be written as 
\begin{equation}
\bm{\theta}=(\alpha, \beta_1 ,..., \beta_k, \gamma)^t,
\label{eq:SIR_varying_betas}
\end{equation}
where $\beta_j$ represents the infection rate at the $j$-th set of time intervals, with $j =1,...,k$, and $k$ represents the total number of different infection rates allowed in the model.
This means that for the $i$-th observation  period, the corresponding SIR process is parameterized by $(\alpha, \beta_i, \gamma)^t$, where some $\beta$s may be constrained to be equal across observation periods.

\subsection{Parameters Inference}
\indent \par 
If we want to proceed in a Bayesian framework, then ideally we would like to approximate the posterior distribution of the SIR parameters $\bm{\theta}$, the true susceptibles $\mathbf{S}$, and true infectious $\mathbf{I}$ subpopulations, given by:
\begin{equation}
p(\bm{\theta},\mathbf{S}, \mathbf{I}|\mathbf{Z})
\propto 
p(\mathbf{Z}|\mathbf{S}, \mathbf{I})
\cdot p(\mathbf{S}, \mathbf{I}|\bm{\theta})
\cdot p(\bm{\theta}),\label{eq:posteriorIdeal}
\end{equation}

\noindent where $\mathbf{S}=(S_1,...,S_n)^t$, $\mathbf{I}=(I_1,...,I_n)^t$ and $\mathbf{Z}$ represents the vector of incidence records. 
However, approximating the posterior distribution (\ref{eq:posteriorIdeal}) is a difficult computational problem since the values for $\mathbf{S}$ and $\mathbf{I}$ are discrete and high dimensional. 

In general, the imputation of $\mathbf{S}$ and $\mathbf{I}$ can not be done  but techniques exist for endemic infectious diseases under certain assumptions. 
In particular, TSIR provides an ingenious method to  estimate $\mathbf{\tilde{S}}=(\tilde{S}_1,...,\tilde{S}_n)^t$ and $\mathbf{\tilde{I}}=(\tilde{I}_1,...,\tilde{I}_n)^t$ from $\mathbf{Z}$. 
This method relies on the assumption that incidence and prevalence are close enough to be considered almost identical at any observation time, since the observation times are approximately equal to the infectious period. 
Also, it is assumed that information about new births in the population is available along with the incidence records.
Following these assumptions, TSIR provides a reconstruction of the susceptible dynamics based on a recursive model that accounts for births, new infections and a reporting probability.  
Then, a regression model is derived between the cumulative sum of births and the cumulative sum of cases.
The slope of this model is interpreted as the reporting rate, $\rho$, from which the true cases are estimated.
The full description of the TSIR model can be found in \citep{finkenstadt2000dynamical}.

After $\tilde{\mathbf{S}}$ and $\tilde{\mathbf{I}}$ are estimated, we approximate the posterior distribution of the SIR model parameters using

\begin{equation}
p(\bm{\theta}|\mathbf{Y})\propto p(\mathbf{Y}|\bm{\theta})\cdot  p(\bm{\theta}),
\label{eq:posterior}
\end{equation}

\noindent where $\mathbf{Y}=(\tilde{\mathbf{S}}, \tilde{\mathbf{I}})$, and $p(\bm{\theta})$  represents the prior distribution over the SIR model parameters.



\subsection{SIR likelihood function}

\indent \par The likelihood function from equation (\ref{eq:posterior}) can be written as the product of the Markov process transition probabilities

\begin{equation}
 p(\mathbf{Y}|\bm{\theta}) = \prod_{t=1}^{n-1} P_{\mathbf{X}_t,\mathbf{X}_{t+1}}(\Delta t),
\label{eq:CTMC_likelihood}
\end{equation}
where the transition probabilities are
\begin{equation}
P_{\mathbf{X}_t,\mathbf{X}_{t+1}}(\Delta  t)=\mathbb{P}[\mathbf{X}(t+\Delta t)=\{\tilde{S}_{t+1},\tilde{I}_{t+1} \} \, | \,
\mathbf{X}(t)=\{\tilde{S}_{t},\tilde{I}_{t} \} ].
\label{eq:transition_probab}
\end{equation}
With the exception of the most simple cases \citep{feller1968introduction,allen2010introduction}, statistical inference involving these transition probabilities can be quite difficult. 
A common approach is to solve the forward Kolmogorov equations by matrix exponentiation techniques. 
However, these techniques can be computationally prohibitive for stochastic systems with a large number of states such as SIR models \citep{keeling2008methods}. 
Instead, we consider two alternative approaches where the exact transition probabilities are calculated: 
using a negative binomial approximation for the infection process which is part of the TSIR formulation, and using a highly accurate numerical approximation technique that explores the fact that the the SIR model can be  represented by a death/birth-death process \citep{ho2018birth,ho2018direct}.

\subsubsection{Negative binomial approximation to the transition probabilities. \label{sec:Methods_NegBin}}

A negative binomial distribution model
has been previously adopted by the TSIR framework to approximate SIR transition probabilities  \citep{grenfell2002tsir,bjornstad2008hazards}.
This model is derived via a linear pure birth approximation of the SIR model, as summarized in \citep{wakefield2019spatio}, and provides a simple expression to calculate the transition probabilities (\ref{eq:transition_probab}).
In this derivation, the law of mass action assumption of homogeneous mixing among susceptible and infectious individuals is required. 
More specifically, one assumes the number of susceptible individuals is approximately constant ($S(t) = s$) during the time period of interest and the number of infectious individuals follows a pure birth process:
\begin{equation}
\mathbb{P}(I(t + \Delta t) = j \mid   I(t)  = i )= 
\begin{cases}
 \frac{ \beta }{N} s i^{\alpha}  + o(\Delta t) & \text{if }  j  = i+1,\\
o(\Delta t) &\text{otherwise}.
\end{cases}
\label{eq:generalized_pure_birth}
\end{equation}
When $\alpha =1$, properties of the \textit{linear}  pure-birth process imply that the number of infectious individuals at time $t$, given the number of susceptible and infectious individuals at $t-1$, can be written as
\begin{equation}
    I_t \mid (S_{t-1},I_{t-1}) 
    \sim NegBin(m_t, I_{t-1}),
    \label{eq:NegBin_likelihood}
\end{equation}
where the mean of the distribution is 
\begin{equation} 
m_t = I_{t-1} \left( e^{\beta S_{t-1} /N} - 1 \right). 
\label{eq:NegBinExp_mean}
\end{equation}
TSIR models use the first order Taylor expansion of equation (\ref{eq:NegBinExp_mean}) and write  the mean of the negative binomial model (\ref{eq:NegBin_likelihood}) as   
\begin{equation}
m_t = \beta \,S_{t-1} I_{t-1} /N .
\label{eq:NegBin_mean}
\end{equation}

Previous literature assumes this model can be generalized to allow for non-homogeneous mixing simply by raising the infectious population to a mixing power, $\alpha$, in the mean of the distribution, $m_t = \beta S_{t-1} I_{t-1}^\alpha /N $ \citep{finkenstadt2000dynamical, grenfell2002tsir}.
This modification can also be applied to equation (\ref{eq:NegBinExp_mean}) to obtain $m_t = I_{t-1}^{\alpha} \left( e^{\beta S_{t-1} /N} - 1 \right)  $). 
However, neither of these approximations are mathematically justified, because the negative binomial approximation of SIR transition probabilities arises  from the linear birth approximation and is valid only when $\alpha=1$. 
This incorrect generalization can lead to model misspecification as we show in Section \ref{sec:Results}.
Also, note that since the negative binomial model was derived using a linear birth approximation, this model should be valid under pure birth conditions, which corresponds to the case when the size of the susceptible population is significantly larger than the number of infectious individuals. 
One example of this setting is the beginning of an epidemic outbreak where the infectious population grows exponentially.
Finally, within the context of TSIR, since the time between observations is assumed to match the infectious period of a disease, the negative binomial model does not consider recovery events; thus, the recovery rate should be approximately one and is not estimated.

\subsubsection{Numerical approximation of the transition probabilities \label{sec:Methods_MultiBD}}
Recently, numerical algorithms  for birth/birth-death processes  has been introduced to efficiently evaluate finite time transition probabilities of these bivariate stochastic processes \citep{ho2018birth,ho2018direct}.
This class of processes contains the SIR model (\ref{eq:Stochastic-SIR}) when it is expressed as a  death/birth-death process (Figure \ref{fig:CompartmentalModel}).
The idea is to reparameterize the SIR process in terms of the cumulative number of infections, $nSI(t)$, and the cumulative number of removals, $nIR(t)$.
The original random variables $S(t)$ and  $I(t)$ can be recovered deterministically from $nSI(t)$, $nIR(t)$, $S(0)$, and $I(0)$.
For the case of a SIR process with this alternative parametrization, the \texttt{MultiBD} package in R provides an efficient and numerically stable implementation for the calculation of the transition probabilities required for direct likelihood based inference.
This means that if we are able to reconstruct the susceptible and infectious populations from incidence records, we can estimate the posterior distribution of the model parameters $\bm{\theta}$  (\ref{eq:posterior}) using standard MCMC methods.
The \texttt{MultiBD} package is able to calculate the required transition probabilities needed to execute an MCMC algorithm.
\par
We also use the  \texttt{MultiBD} exact method to compute transition probabilities of the non-linear pure birth process \eqref{eq:generalized_pure_birth}. 
In this pure birth regime, the \texttt{MultiBD} exact method matches the Negative Binomial probabilities from Equation \eqref{eq:NegBin_likelihood} when $\alpha =1$ in \eqref{eq:generalized_pure_birth}. 
However,  the negative binomial model returns incorrect transition probabilities when the mixing rate differs from unity as we show in Section \ref{sec:Results}.

\subsection{Methods Summary}

Starting from incidence data, we follow the TSIR approach to reconstruct the time series of susceptible and true cases. 
Then, to evaluate the likelihood function (equation (\ref{eq:CTMC_likelihood})), we compute the required transition probabilities (equation (\ref{eq:transition_probab})).
In particular, for BayesTSIR, we obtain transition probabilities using the negative binomial distribution defined by equations (\ref{eq:NegBin_likelihood}) and (\ref{eq:NegBin_mean}).
We also compute transition probabilities of the non-linear pure birth process \eqref{eq:generalized_pure_birth}.
When we use these transition probabilities in Bayesian inference, we call the corresponding statistical method \textit{BayesPureBirth}.
For \textit{BayesSIR}, we compute the transition probabilities using the exact method by \citet{ho2018birth,ho2018direct}. 
Finally, using the three Bayesian inference approaches, we approximate the posterior distribution of the model parameters (\ref{eq:posterior}) using MCMC methods.
We summarize all transition probability approximations and the corresponding names of inferential procedures  in Table \ref{tab:summary_methods}.
The source code for the computational implementations in this paper can be found at \url{https://github.com/luisdm1/BayesSIR}. Also, a slight  extension of the \texttt{MultiBD} package to compute pure birth process transition probabilities can be found at \url{https://github.com/vnminin/MultiBDPureBirth2}. 

\begin{table}
\caption{Summary of statistical methods. We show the method names, the transition probabilities method, computational implementation and reference to the section in the main text where they are introduced. }
\centering
\begin{tabular}{cccc}
Method & Transition Probabilities & Description & Section\tabularnewline
\toprule 
\textit{BayesSIR} & MultiBD-SIR & Exact method & \ref{sec:Methods_MultiBD}\\
\textit{BayesPureBirth} & MultiBD-PureBirth  & Exact method & \ref{sec:Methods_MultiBD}\\
\textit{BayesTSIR} & NegBin & Negative binomial & \ref{sec:Methods_NegBin}\\
NA & NegBinExp & Exponential negative binomial &\ref{sec:Methods_NegBin} \\
\end{tabular}
\label{tab:summary_methods}
\end{table}

\section{Results\label{sec:Results}}
\indent \par In this section, we first show conditions under which transition probabilities under different approximations take on similar values. 
We then provide insights about when these  approximations diverge from each other and from exact calculations.
Next, we compare the performance of our Bayesian implementation of the TSIR negative binomial model (\textit{BayesTSIR}) and the SIR model with constant and time-varying infection rates (\textit{BayesSIR}) for estimating mixing, infection, and removal rates.
We apply our method to simulated prevalence data, and to a benchmark dataset from historic incidence measles records. 
When analyzing simulated data, we also include results of the \textit{BayesPureBirt}h method to unambiguously demonstrate that 
differences between \textit{BayesTSIR} and \textit{BayesSIR} methods cannot  be explained  by the  pure birth  approximation alone.
Our results show that using \textit{BayesTSIR} model results in potential bias in the parameter estimates and suggests exaggerated fluctuations of the infection rate estimates across time.

\subsection{Transition probabilities comparison}

\begin{figure}
    \includegraphics[width=\textwidth]{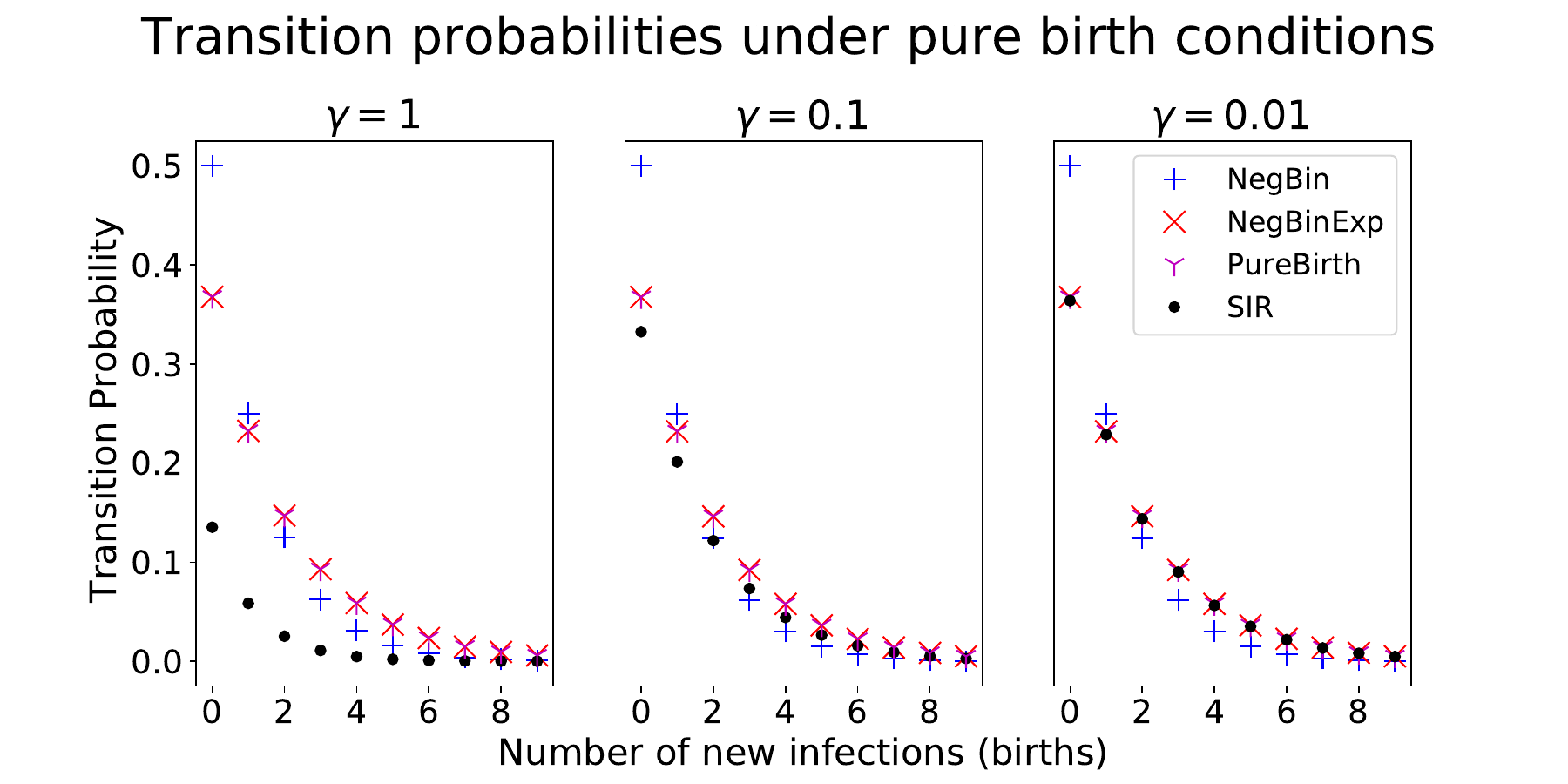}
  \centering
  \caption{
  Comparison of the transition probabilities (equation (\ref{eq:transition_probab})) using the two formulations of the negative binomial model (equations (\ref{eq:NegBinExp_mean}),  (\ref{eq:NegBin_mean})), a pure birth process and a SIR process when the recovery rate, $\gamma$, is decreased.
  Starting from an initial condition with ($S_0=999999$, $I_0=1$, $R_0=0$), the plots show the probability of obtaining $i\in\{0,1,...,9\}$ new infections (interpreted as new births in the linear birth approximation  from TSIR).
  We assume a time step $\Delta t=1$, homogeneous mixing, $\alpha=1$,  constant infection rate, $\beta=1$.
  Also, we assume no recoveries, $nIR=0$, in the MultiBD parametrization of the SIR process and in the pure birth approximation. 
  }
  \label{fig:Trans_prob_comparison}
\end{figure}

\begin{figure}
    \centering
    \includegraphics[width=0.95\textwidth]{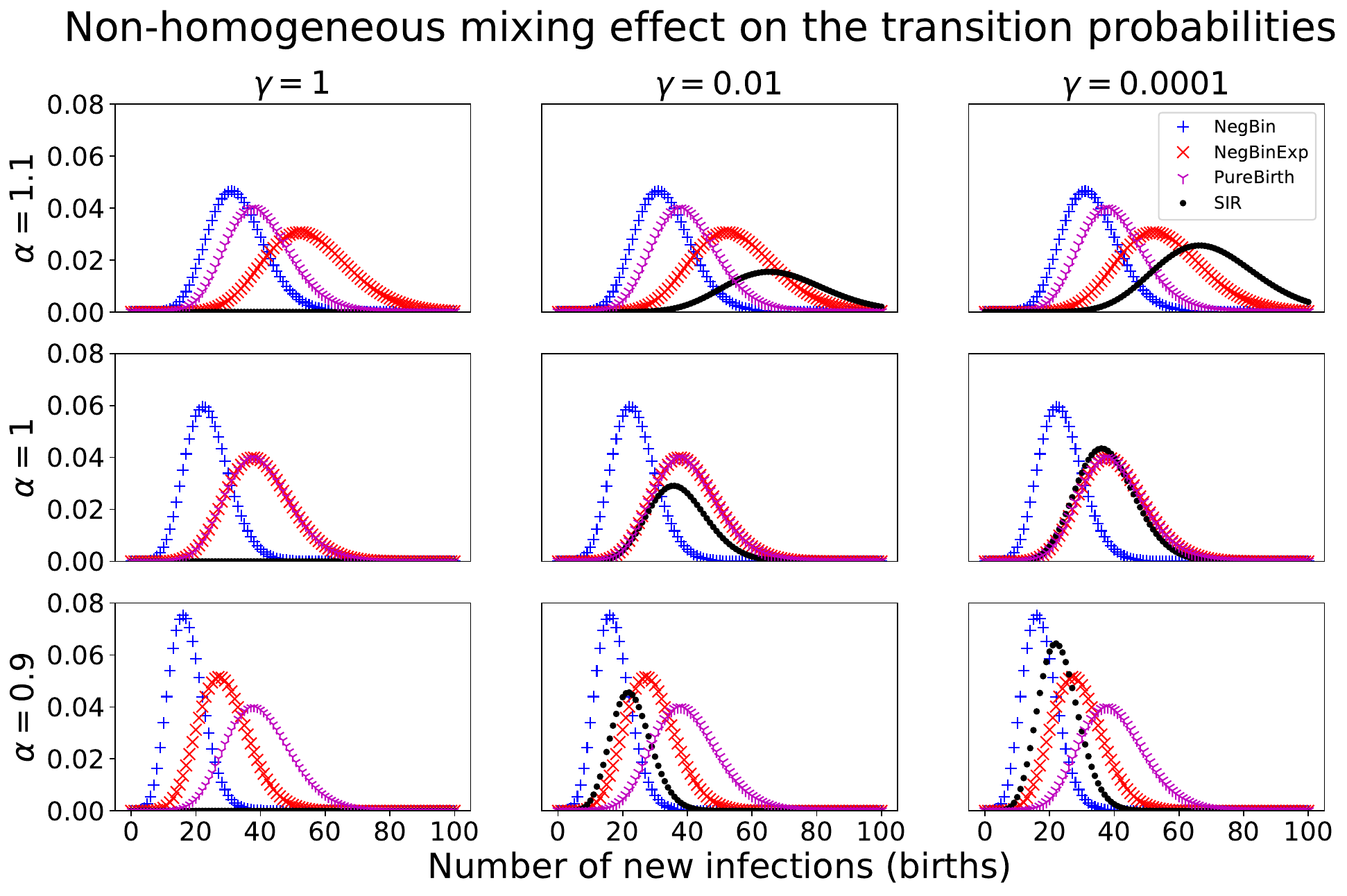}
    \label{fig:Trans_prob_saturation}
    \caption{Comparison of the transition probabilities (equation (\ref{eq:transition_probab})) using the two formulations of the negative binomial model (equations (\ref{eq:NegBinExp_mean}),  (\ref{eq:NegBin_mean})), a pure birth process and a SIR process.
    Here, we show the effect of non-homogeneous mixing on the transition probabilities by decreasing the mixing rate $\alpha\in\{1.1, 1, 0.9 \}$ and the recovery rate, $\gamma\in\{1, 0.01, 0.0001\}$.
    The time step is $\Delta t=1$, the number of recoveries is $nIR=0$, the infectious rate is set to $\beta=1$, and the initial conditions for the SIR compartments are ($S_0=500$, $I_0=25$, $R_0=0$).
    }
    \label{fig:Trans_prob_non_homog}
\end{figure}

\indent \par Our approach to the likelihood-based inference for prevalence and incidence data relies on the calculation of the transition probabilities shown in equation (\ref{eq:transition_probab}).
However, model misspecification can lead to incorrect calculation of the these transition probabilities and severely impact the estimation of infection, removal and mixing rates.
This can occur under the negative binomial model because it only considers new infections (births) but no removals.
Therefore, the negative binomial model should be valid under a pure birth regime when the number of susceptible individuals does not change substantially and when the number of removals is negligible. 
For instance, these conditions occur at the beginning of an epidemic outbreak where there is an exponential growth phase of the number of new infections and the time period of interest is short.
In this regime, the NegBin (\textit{BayesTSIR}), PureBirth (\textit{BayesPureBirth}), and SIR (\textit{BayesSIR}) methods produce  similar transition probabilities as observed in the right plot of Figure \ref{fig:Trans_prob_comparison}.
This is due to the fact that new infections are more likely to occur than removals when the population of susceptible individuals is larger than the number of infectious cases, which induces a small  effective removal rate $\gamma I$. 
Note that this can also occur when the removal rate $\gamma$ is small.
In contrast, when the recovery rate is large, there is a mismatch between the transition probabilities obtained using the exact \texttt{MultiBD} method and pure birth approximations since new removals are now more likely to occur. 
Since in these numerical experiments we assume that $\alpha  =1$,  PureBirth and NegBinExp transition probabilities coincide exactly, as predicted by theory.
Moreover, these distributions do not change when varying the removal rate since, by definition, these methods do not model this type of  events as explained in Section \ref{sec:Methods}. 

Next, we compare transition probability calculation methods when varying both removal rate $\gamma$ and mixing rate $\alpha$. 
When we consider non-homogeneous mixing ($\alpha \ne 1$) under a pure birth regime, the NegBin and SIR transition probabilities are relatively close to each other  when the recovery rate is small as shown in Figure \ref{fig:Trans_prob_non_homog}. 
However, as $\gamma$ increases, there is a significant mismatch between the NegBin and SIR transition probabilities.
For the pure birth models, we see that NegBinExp and PureBirth show equivalent transition probabilities during homogeneous mixing only. 
This shows that negative binomial models do not adequately approximate non-homogeneous mixing even under the pure  birth  regime.
Finally, we observe in Figure \ref{fig:Trans_prob_non_homog} the mechanistic effect of the mixing rate on the transition probabilities distribution, i.e. this parameter can accelerate or delay the infection dynamics due to its non-linear effect on the rate of accumulation of new infections given by $\beta \,S\, I^\alpha /N$.
As a result, the distribution of the transition probabilities can be shifted lower or higher  when $\alpha<1$ or $\alpha>1$ respectively.

\subsection{Inference using prevalence data}

\indent \par  When prevalence data is available and sizes of the susceptible and removal compartments are known,
it is possible to estimate the parameters of a SIR model using the direct likelihood calculations afforded by the analytical Negative Binomial  formula under the BayesTSIR model and exact transition probability calculations under BayesPureBirth and BayesSIR models (Section \ref{sec:Methods}).
To illustrate this approach, we simulate stochastic SIR trajectories using the Gillespie algorithm and fixing the initial conditions to $(S_0=999,I_0=1,R_0=0)$ and the SIR model parameters $\beta=0.0045$ and  $\gamma=1$.
The effect of non-homogeneous mixing is implemented by setting $\alpha=0.8$.

\begin{figure}
    \centering
     \includegraphics[width=\textwidth]{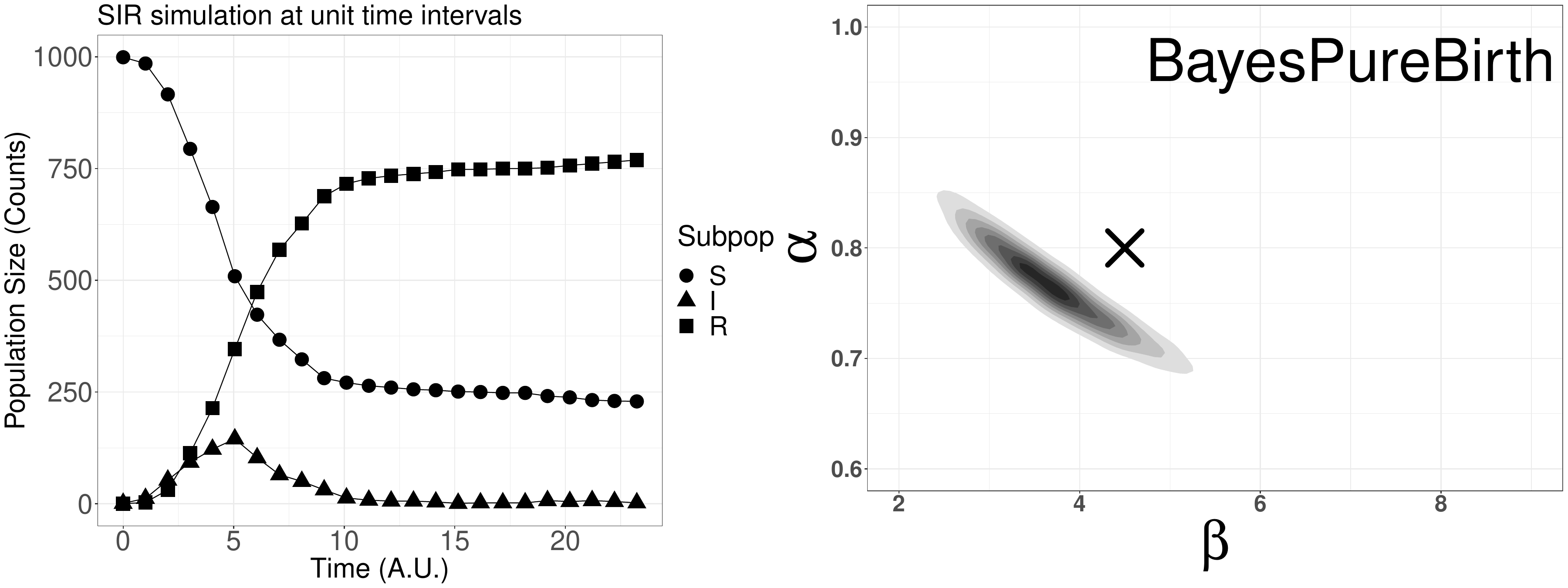}
  \includegraphics[width=1.0\textwidth]{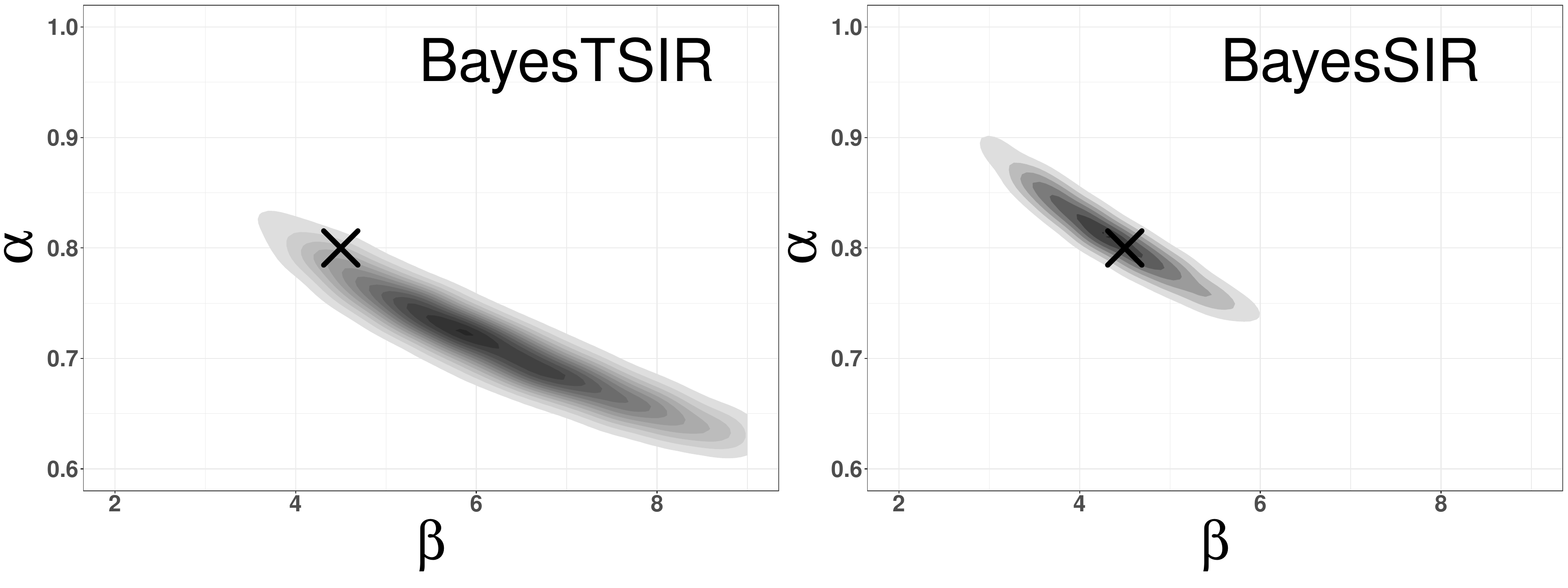}
  \centering
  \caption{\label{fig:Prev_inference_plots}
    Statistical inference for simulated prevalence data under non-homogeneous mixing.
      \textbf{Top Left}: Example of simulated stochastic SIR trajectories with initial conditions $S_0=999$, $I_0=1$ and parameters  $\alpha=0.8$, $\beta=0.0045$, $\gamma=1$. 
      \textbf{Remaining subfigures}: Joint posterior distributions of mixing ($\alpha$) and infection ($\beta$) rates obtained with BayesPureBirth, BayesTSIR and BayesSIR methods.
      For BayesSIR, the posterior median for $\gamma$ was estimated to be 1.045 with a 95\% Bayesian credible interval (0.9693, 1.128). Note that this parameter was estimated by neither BayesPureBirth nor BayesTSIR.
      The cross designates the true parameter values used in the simulation.
  }
  \end{figure}

We show in Figure \ref{fig:Prev_inference_plots} the joint posterior distributions of the extended SIR mixing parameter ($\alpha$) and infection rate ($\beta$) obtained by analyzing one simulated data set shown in the same figure. 
As expected, the BayesSIR method successfully recovers the true parameters values.
Interestingly, the BayesPureBirth method produces a posterior distribution that is similar in shape to the BayesSIR  distribution, but shifted away from the true values. 
Nonetheless, the marginal posterior distributions of $\alpha$) and $\beta$ look reasonable even in this example.
BayesTSIR  posterior is centered far from the true parameters, but its large spread along the $\beta$ axis allows it capute the true parameter values in the tail of the distribution. 


To investigate frequentist properties of the above Bayesian methods, we performed a simulation study by generating an ensemble of 629 SIR simulations with fixed parameters under inhomogeneous mixing. 
At each iteration, a synthetic SIR trajectory was generated using the Gillespie algorithm until the depletion of the infectious compartment and values of the resulting SIR trajectories were recorded at a regular grid of time points.
For each simulated  trajectory,  we performed a MCMC simulation to approximate the posterior distribution of the model parameters for each of the three methods: BayesSIR, BayesPureBirth and BayesTSIR.
We report root mean squared errors (RMSEs), mean widths of  95\% Bayesian credible intervals (BCIc), and coverage of 80\%, 90\%, and 95\% BCIs in Table~\ref{tab:coverage_table}.
BayesPureBirth  and BayesSIR have similar RMSEs and BCI widths for $\alpha$ and $\beta$.
However, only BayesSIR BCI coverages are close to their nominal values.
BayesPureBirth BCI coverages are significantly lower.
BayesTSIR produces the highest root mean squared errors, widest 95\% BCIs and the lowest coverages of all types of BICs.
Note that only BayesSIR estimates the recovery rate and does it well, as shown in the last row of Table~\ref{tab:coverage_table}.
Our simulation results show that a pure birth approximation to the SIR with inhomogeneous mixing leads to bias and poor frequentist calibration of BCIs. 
Further, using the negative binomial distribution exaggerates these problems further and in addition reduces estimation precision.

\begin{table}
\caption{Simulation study results. 
We generated 629 trajectories of a SIR stochastic process using parameters $\alpha=0.8$, $\beta=0.0045$, and $\gamma=1$, similarly  to the  simulation shown in Figure \ref{fig:Prev_inference_plots}. Then, we used BayesTSIR, BayesPureBirth and BayesSIR and likelihood formulations  to approximate their corresponding posterior distributions of these parameters using MCMC. For each parameter and method specified, we report the Root Mean Squared Errors (RMSEs) of the posterior median, mean widths of the 95\% Bayesian credible intervals (BCIs), and the coverage proportions for BCIs at the 80\%, 90\% and 95\% credible levels.}
\centering
\begin{tabular}{ccccccc}
& & & Mean width & \multicolumn{3}{c}{\fontsize{12}{18}\selectfont
Coverage}\\

Parameter &  Method & RMSE & 95\% BCI & 80\% & 90 \% & 95\% \\ \toprule

 & BayesTSIR & $0.13$&$0.25$ & $0.23$&$0.35$&$0.49$\\

\fontsize{18}{18}\selectfont $\bm{\alpha}$ & BayesPureBirth & $0.07$ & $0.16$ & $0.45$& $0.57$ & $0.66$\\

 & BayesSIR  & $0.05$ & $0.18$&$0.76$&$0.87$&$0.92$ \\ \midrule

 & BayesTSIR & $3.73$&$8.69$ & $0.23$&$0.38$&$0.49$\\

\fontsize{18}{18}\selectfont $\bm{\beta}$ & BayesPureBirth & $0.89$&$3.02$ & $0.69$ & $0.79$ & $0.84$ \\

 & BayesSIR  & $1.06$ & $3.83$ & $0.77$ & $0.87$ & $0.93$ \\ \midrule
\fontsize{18}{18}\selectfont $\bm{\gamma}$ & BayesSIR & $0.04$&$0.15$ & $0.78$& $0.90$ & $0.95$\\ 

\end{tabular}
\label{tab:coverage_table}
\end{table}

\subsection{London Measles Case Study \label{sec:DataAnalysis}}

\indent \par We apply our method to the classic measles dataset studied by \cite{finkenstadt2000dynamical}. 
This dataset contains biweekly incidence reports of measles cases between 1944 and 1964. 
The time series shows the cyclic pattern of measles transmission across years, which was hypothesized to  be related to the school calendar.
In particular, previous studies have identified lower infection rates during school breaks compared to school terms \citep{klinkenberg2018reduction}. 
We are interested in revisiting this question when comparing BayesSIR and BayesTSIR.

\begin{figure}
	\centering
	\includegraphics[width=0.9\textwidth]{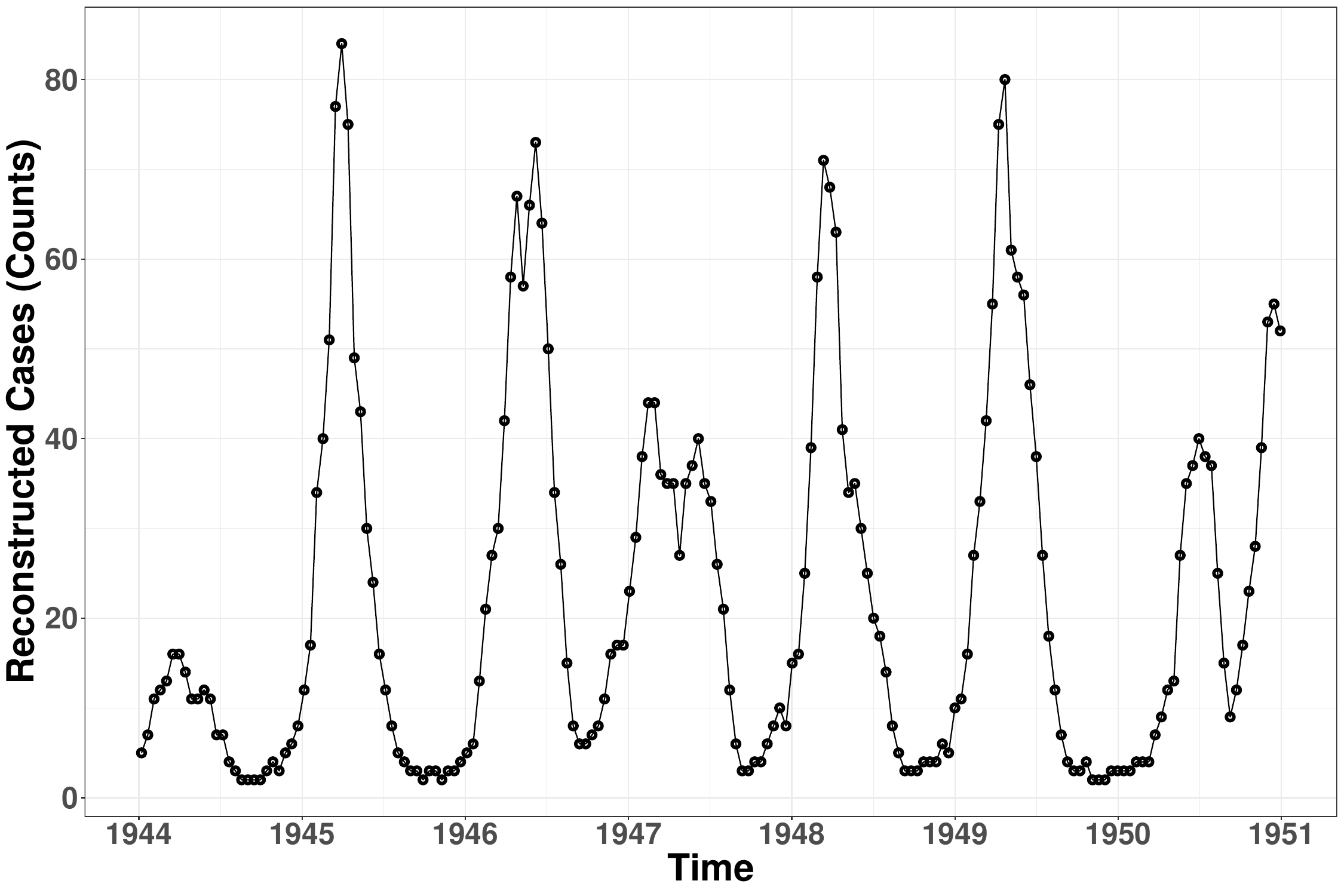}
   \caption{\label{fig:LondonData} Reconstructed London measles dataset from 1944 to 1951 after applying a rescaling factor of 100 to the original time series.}
\end{figure}

Our approach requires computation of the transition probabilities to obtain the likelihood function (\ref{eq:CTMC_likelihood}).
However, computing these exact SIR transition probabilities using MultiBD for the BayesSIR method is problematic since high SIR compartment counts (on the order of millions) make repeated evaluation of transition probabilities during MCMC computationally prohibitive. 
To make this problem tractable, we first take a subset of the data by only considering years 1944--1951.
Next, we divide the incidence cases by 100 and round them to the nearest integer as shown in Figure \ref{fig:LondonData}.
This down-scaling is reasonable since we still preserve the same measles dynamics, including  seasonality, but artificially reduce the size of the population under surveillance. 

The first step in our analysis is the estimation of the numbers of susceptible and latent infectious individuals.
For this purpose, we use the tSIR package \citep{becker2017tsir} to estimate the susceptible population ($\tilde{\mathbf{S}}$), the reporting probability ($\rho$), and the true number of  infections ($\tilde{\mathbf{I}}$) from the incidence records shown in Figure \ref{fig:LondonData}.

Following previous analyses of this dataset, we assume that the infection rates are not identical throughout the year but the mixing and removal rates remain constant. 
In particular, we consider two types of seasonality: \textit{schoolterm}, which depends on whether schools are in session or not (corresponding to two infectious rates) and \textit{standard}, which consists of 26 biweeks in a typical year (corresponding to one infectious rate per biweek).
For the standard seasonality model, TSIR is able to estimate 26 infectious rates (following the generalized linear model implementation in the R package \texttt{tSIR}).
However, due to computational challenges for estimating the transition probabilities using MultiBD, we only computed 13 infectious rates which corresponds to roughly one infectious rate every four weeks.
For consistency, we parameterized  all methods to include 13 infectious rates.


We use the BayesSIR method  to estimate the infection, mixing, and removal rates for the measles dataset.
Assuming the schoolterm seasonality model, we estimate two infection rates $\bm{\theta}_{schoolterm}=$ $(\alpha, \beta_{shoolbreak},$ $\beta_{schoolterm},\gamma)^t$.
Similarly, for the standard seasonality model, we estimate the parameters $\bm{\theta}_{standard}=$ $(\alpha,$ $\beta_1,$ $\beta_2,$ $ \dots,$ $\beta_{13},$  $\gamma)^t$. 
We use the BayesTSIR with the same settings, except this method does estimate the removal rate $\gamma$  and sets it to 1.0 instead, because the inter-observation time interval length roughly matches measles mean infectious period of two weeks.
We use the Metropolis-Hastings algorithm to approximate the posterior distribution (\ref{eq:posterior}) for both models where we use independent $\textrm{LogNormal}(0, 100^2)$ priors for all the parameters.

\begin{table}
\caption{Estimated mixing ($\alpha$) and recovery ($\gamma$) rates for the subsampled London  measles dataset.
We report the posterior median and the 95\% Bayesian credible intervals (BCIs) for the BayesSIR and BayesTSIR methods. We also report  TSIR  point estimates ($^*$) of these parameters calculated using the \texttt{tsiR} R package \citep{becker2017tsir}.}
\centering
\begin{tabular}{cccccc}
& & \multicolumn{2}{c}{\fontsize{12}{18}\selectfont
Mixing rate $\alpha$ } & \multicolumn{2}{c}{\fontsize{12}{18}\selectfont
Recovery rate $\gamma$}\\

Seasonality &  Method & Median & 95\% BCI & Median & 95\% BCI\\\midrule

 & BayesSIR & $0.965$&$(0.922,1.01)$ & $0.99$&$(0.98,1.02)$\\

Schoolterm & BayesTSIR & $0.945$&$(0.888,1.00)$ & --& --\\

 & TSIR  & $0.998^*$ & --&-- &-- \\ \midrule

 & BayesSIR & $0.915$&$(0.861,0.973)$ & $0.991$&$(0.959,1.026)$\\

Standard & BayesTSIR & $0.941$&$(0.866,1.014)$ & -- & -- \\

 & TSIR  & $0.974^*$ & -- & -- & -- 

\end{tabular}
\label{tab:alpha_power_post_summaries}
\end{table}

\begin{figure}
	\centering
	\includegraphics[scale=0.5]{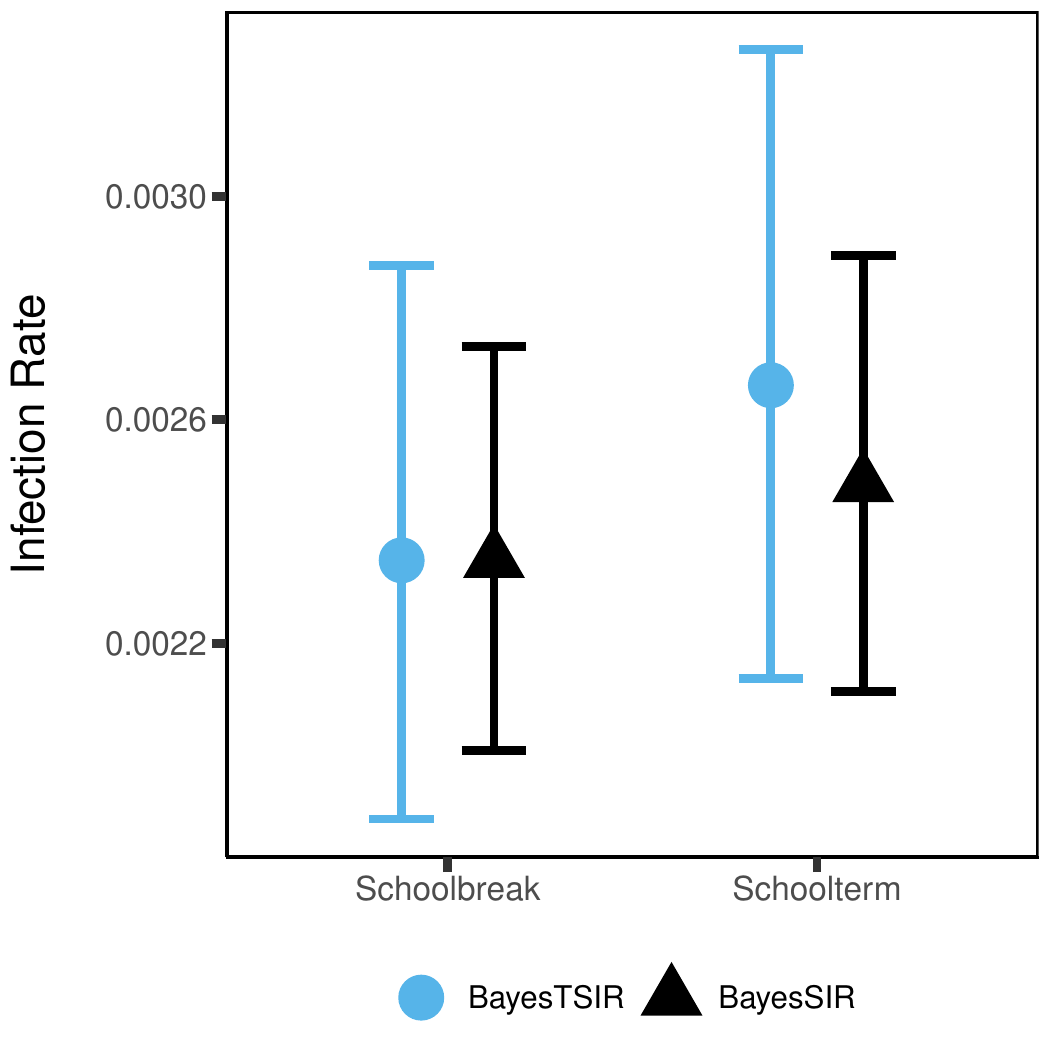}

   \caption{\label{fig:schoolterm} Posterior means (circles and diamonds) and 95\% posterior credible intervals (vertical lines with whiskers) for school term and school break infection rates estimated by BayesSIR and BayesTSIR  methods from the subsampled  London measles data  under the \textit{schoolterm} seasonality  model.}
\end{figure}

Figure \ref{fig:schoolterm} shows the estimated 95\% credible intervals for the infection rates of BayesTSIR and BayesSIR under the \textit{schoolterm} model of  seasonality.
The BayesTSIR exaggerates the decrease in the infection rate during school  breaks compared to BayesSIR. 
Moreover, BayesTSIR results in wider credible intervals for both infection and mixing rates compared to the BayesSIR estimates (Figure \ref{fig:schoolterm} and Table \ref{tab:alpha_power_post_summaries}).
We include in Figure \ref{fig:schoolterm_supp} the infection rate 95\% confidence intervals produced by  the TSIR method as a reference, where we observe a similar pattern.
Note that the BayesSIR  estimates the removal rate are approximately 1.0 as shown in Table \ref{tab:alpha_power_post_summaries}.

\begin{figure}
    \includegraphics[width=0.95\textwidth]{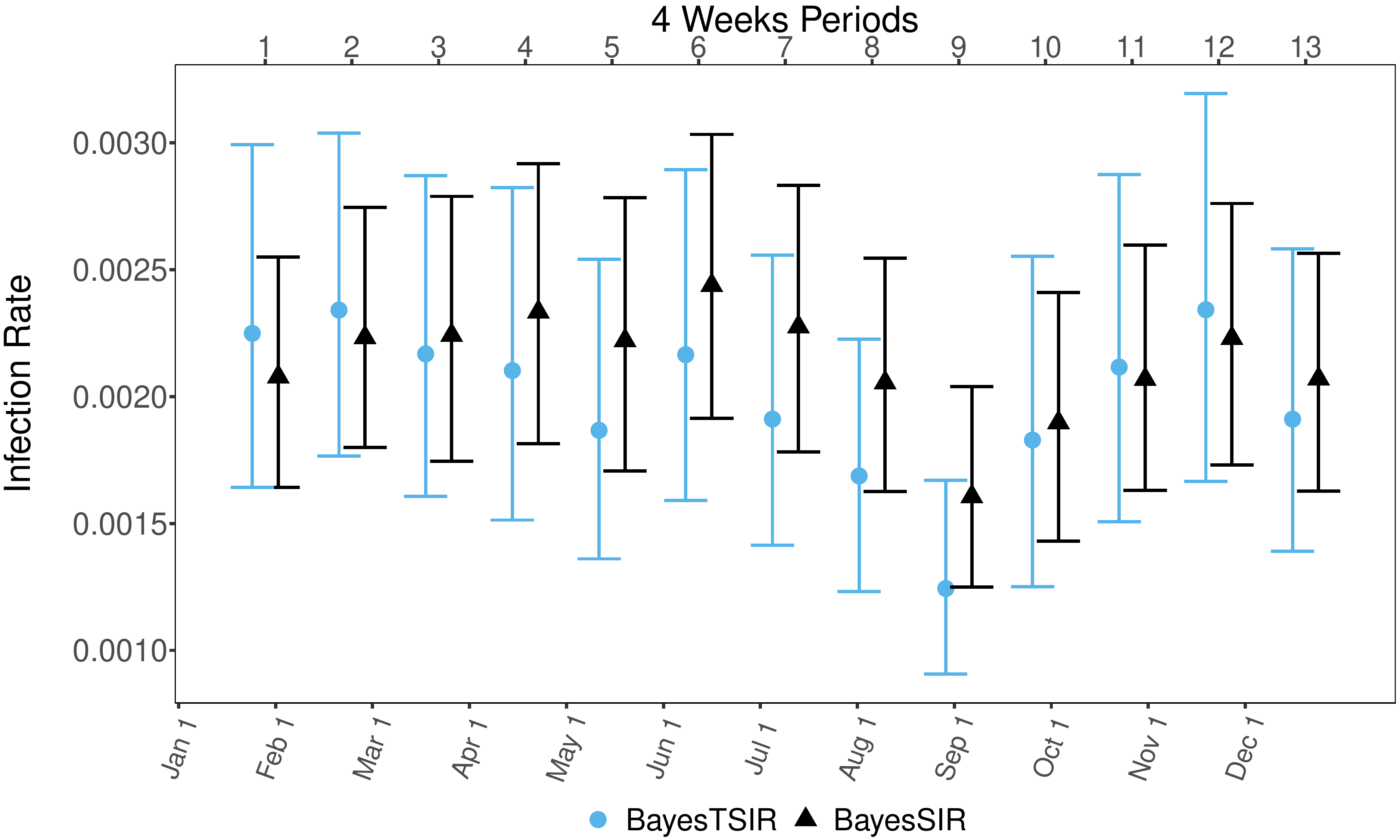}
  \caption{\label{fig:standard} Posterior means (circles and diamonds) and 95\% posterior credible intervals (vertical lines with whiskers) for infection rates estimated by BayesSIR and BayesTSIR  methods from the subsampled  London measles data under the \textit{standard} seasonality  model.}
\end{figure}

When we consider the \textit{standard} seasonality model, we observe that BayesTSIR once again exaggerates the decrease in the infection rate during a typical academic summer break (July--September) when compared to the BayesSIR estimates as shown in Figure \ref{fig:standard}.
Figure \ref{fig:standard_supp} shows that the frequentist TSIR model  exaggerates the school break effect on measles transmission even  more.
Table \ref{tab:alpha_power_post_summaries} indicates that BayesSIR provides stronger evidence in favor of non-homogeneous mixing under the standard seasonality model than its BayesTSIR counterpart, whose
95\% Bayesian credible interval for the  mixing rate $\alpha$ contains 1.0. 

\section*{Discussion \label{sec:Discussion}}

\indent \par In this paper, we study the statistical implications of relaxing the non-homogeneous mixing assumption when estimating infection and mixing parameters from incidence time series data with the help of the TSIR model. 
The results show that the transition probabilities from the TSIR negative binomial model and those from a stochastic SIR model are close to each other only in the homogeneous mixing regime ($\alpha   =1$). 
In the presence of non-homogeneous  mixing ($\alpha<1$) using the negative binomial model results in an intentional and undesirable  model misspecification.
Using simulated data, we demonstrate that this model misspecification results in bias and incorrect quantification of uncertainty in the infection and mixing rates.
In contrast, by using either accurate numerical approximations of the correct SIR transition probabilities, or to a lesser extent the non-linear pure birth approximation of the SIR model, we obtain more accurate estimation of the model parameters and better quantification of uncertainty under  non-homogeneous mixing.
An interesting avenue for future research would be investigating the effect of TSIR negative binomial approximation on forecasting accuracy and calibration under this class of models.

We use TSIR and MultiBD likelihood approximations in a Bayesian framework to estimate the infection, mixing and recovery rates from a subset of the historic London  measles dataset.
Our results show that \textit{BayesTSIR} estimates an exaggerated decrease in the infection rate during academic school break periods compared to \textit{BayesSIR}. 
However, \textit{BayesSIR} reliance on the \texttt{MultiBD} packages makes this method computationally intensive, and currently, it can only be applied to datasets with a moderate number of incidence cases (at most hundreds).
In contrast, the negative binomial likelihood from the \textit{BayesTSIR} is computationally efficient due to the simple analytical form of the transition probabilities.

\par
In conclusion, we showed that results of TSIR models with non-homogeneous mixing should be interpreted  with caution. 
It is tempting to assume that these models approximate the canonical stochastic SIR models with non-homogeneous mixing, but it is not the case.
A non-linear pure birth process would be a valid approximation, but this approximation does not buy us much  in computational efficiency, because its corresponding transition probabilities are not available in closed form  and need to be computed numerically.
When analyzing data from small populations using models  with not too many compartments, accurate numerical methods for computing  their corresponding transition probabilities,  implemented in the  \texttt{MulltiBD} package, are attractive.
When analyzing surveillance data collected from large populations, we recommend resorting to either particle filter algorithms \citep{andrieu2010particle,ionides2015inference} or to principled approximations of stochastic epidemic models \citep{fearnhead2014inference, fintzi2021linear}.

\section*{Acknowledgments}
We  are grateful to Jason Xu and Lam Ho for their help with the \texttt{MultiBD}  R package. 
B.S., L.D.J.M.L, and  V.N.M. were supported by the NSF grant DMS 1936833.

\bibliographystyle{Apalike-JASA}
\bibliography{library}

\newpage

\clearpage

\setcounter{page}{1}
\setcounter{table}{0}
\renewcommand{\thetable}{S-\arabic{table}}
\renewcommand{\thefigure}{S-\arabic{figure}}
\renewcommand{\thesection}{S-\arabic{section}}

\renewcommand{\theequation}{S-\arabic{equation}}
\setcounter{equation}{0}
\setcounter{section}{0}
\setcounter{figure}{0}

\section*{Supplementary Material}

\begin{figure}[ht] 
	\centering
	\begin{tikzpicture}[
	> = stealth, 
	shorten > = 3pt, 
    shorten < = 3pt, 
	auto,
	node distance = 3cm, 
	semithick 
	]

	\tikzstyle{every state}=[
	draw = black,
	thick,
	fill = white
	]

	\tikzset{cross/.style={cross out, draw=black, fill=none, inner sep=0pt, outer sep=0pt}, cross/.default={2pt}}

	\node[state, minimum size = 1cm] (S) {$S$};
	\node[state, minimum size = 1cm] (I) [right of=S] {$I$};

	\draw [red, dashed] (-0.5,-0.5) -- (0.5, -1.5); 
	\draw [red, dashed] (-0.5,-1.5) -- (0.5, -0.5);

	\draw [->] (S) to[bend left] node[auto] {$\beta_{SIR}$} (I);

	\coordinate[right of=I] (I1);
	\draw [->] (I1) to[left] node[above] {$\lambda^{(2)}$} (I);
	\draw [red, dashed] (4.5,-0.5) -- (3.5, 0.5); 
	\draw [red, dashed] (4.5,0.5) -- (3.5, -0.5);

	\coordinate[below of=S] (S2);
	\draw [shorten >=1cm,->] (S) to[below] node[auto] {$\mu^{(1)}$} (S2);

	\node[state, minimum size = 1cm, dashed] (R) [below of =I] {$R$};
	\draw [->] (I) to[below] node[auto] {$\gamma_{SIR}^{I}$} (R);

	\end{tikzpicture}
	\caption{Graphical representation of a SIR compartmental model as a birth/birth-death process for a time step $(t,\; t + \Delta t)$, where $\lambda^{(2)}$'s is the birth rate, $\mu^{(1)}$ is the death rate, $\beta_{SIR}$ is the infection rate and $\gamma_{SIR}^{I}$ is the recovery rate. 
	In this formulation, a susceptible individual can either stay susceptible or become infected, and an infectious individual can either stay infected or become recovered. 
	This representation of a SIR model allows us to re-parameterize it in terms of new infections $nSI$ and new recoveries $nIR$ and associated transition probabilities using MultiBD.} \label{fig:CompartmentalModel}
\end{figure}
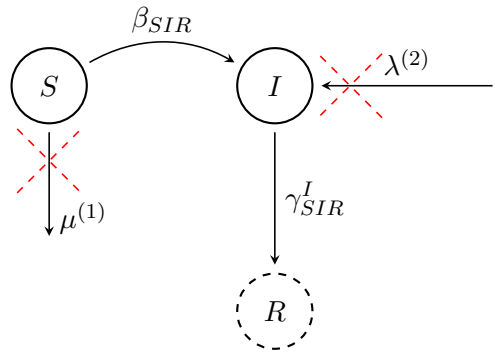

\begin{figure}[ht]
	\centering
	\includegraphics[scale=0.5]{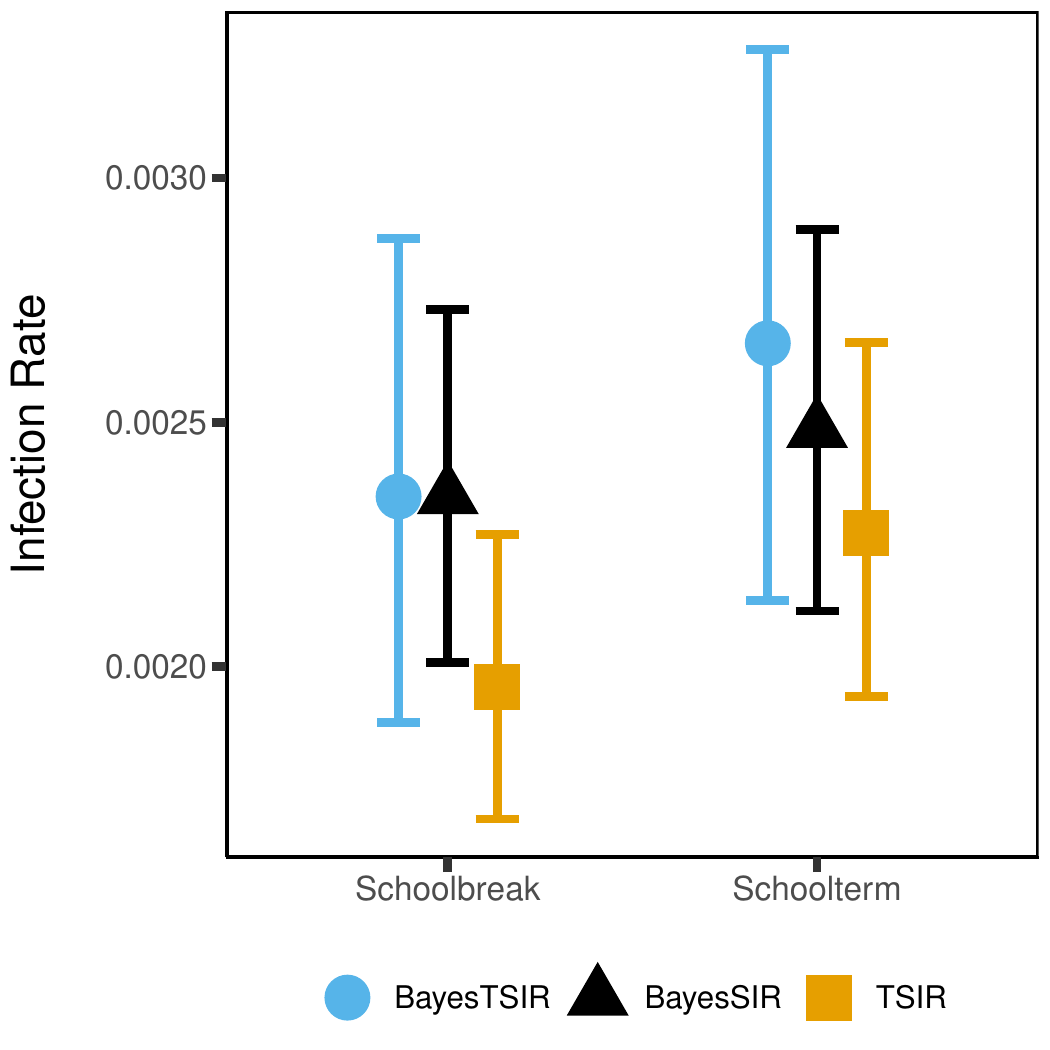}

   \caption{\label{fig:schoolterm_supp} Estimated infectious rates means (marks) and 95\% credible intervals (errorbars) for schoolterm vs. schoolbreak seasonality calculated using the Bayesian TSIR negative binomial model (BayesTSIR) and the Bayesian SIR model with MultiBD (BayesSIR) for the measles dataset from 1944 to 1951. 
   The TSIR 95\% confidence intervals from the tSIR package are included for reference.}
\end{figure}

\begin{figure}[ht]
    \includegraphics[width=\textwidth]{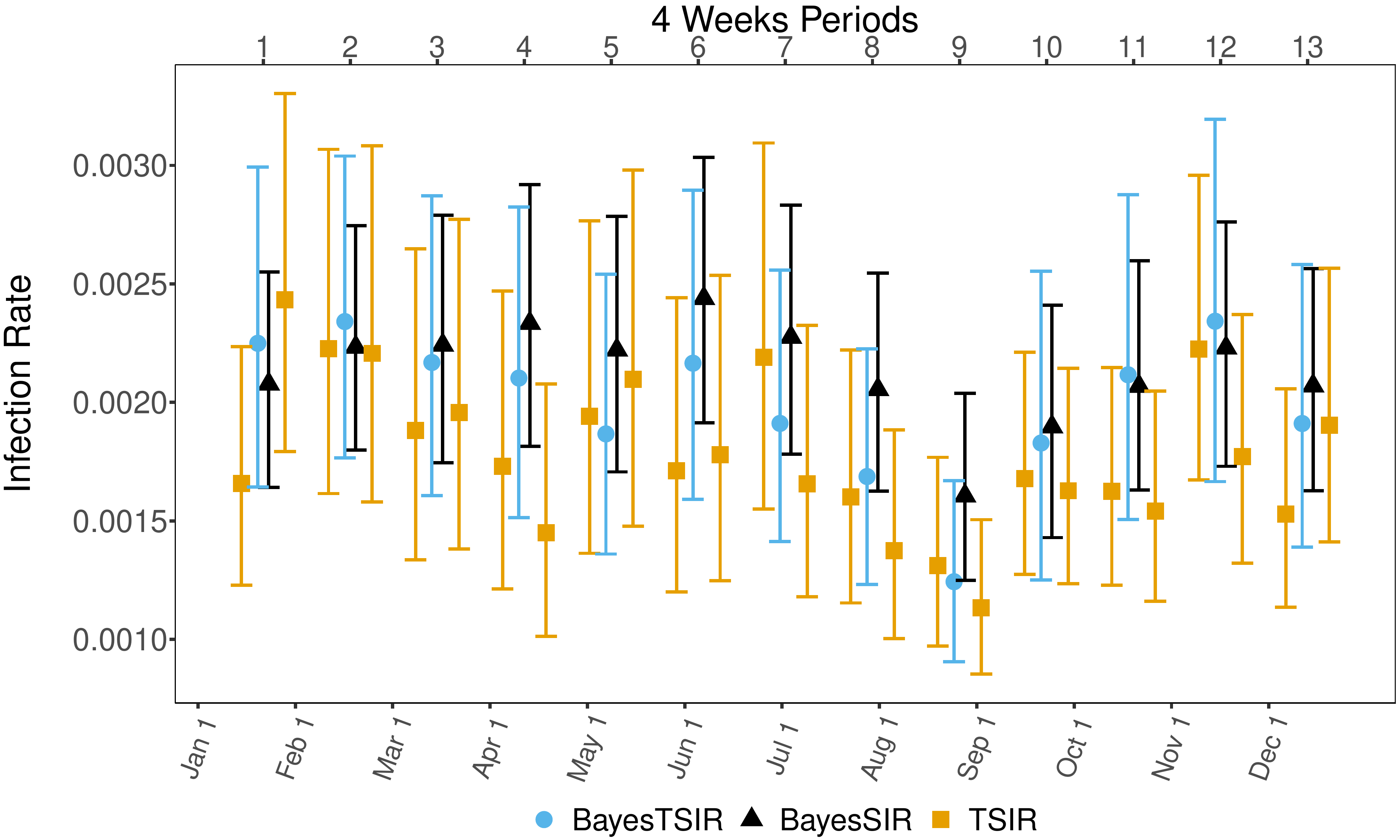}
  \caption{\label{fig:standard_supp} Estimated infectious rates means (marks) and 95\% credible intervals (errorbars) shown every 4 weeks periods calculated using the Bayesian TSIR negative binomial model (BayesTSIR) and the Bayesian SIR model with MultiBD (BayesSIR) for the measles dataset from 1944 to 1951. 
  The TSIR 95\% confidence intervals from the tSIR package are included for reference.}
\end{figure}

\begin{figure}[ht]
\centering

  \includegraphics[width=1\linewidth]{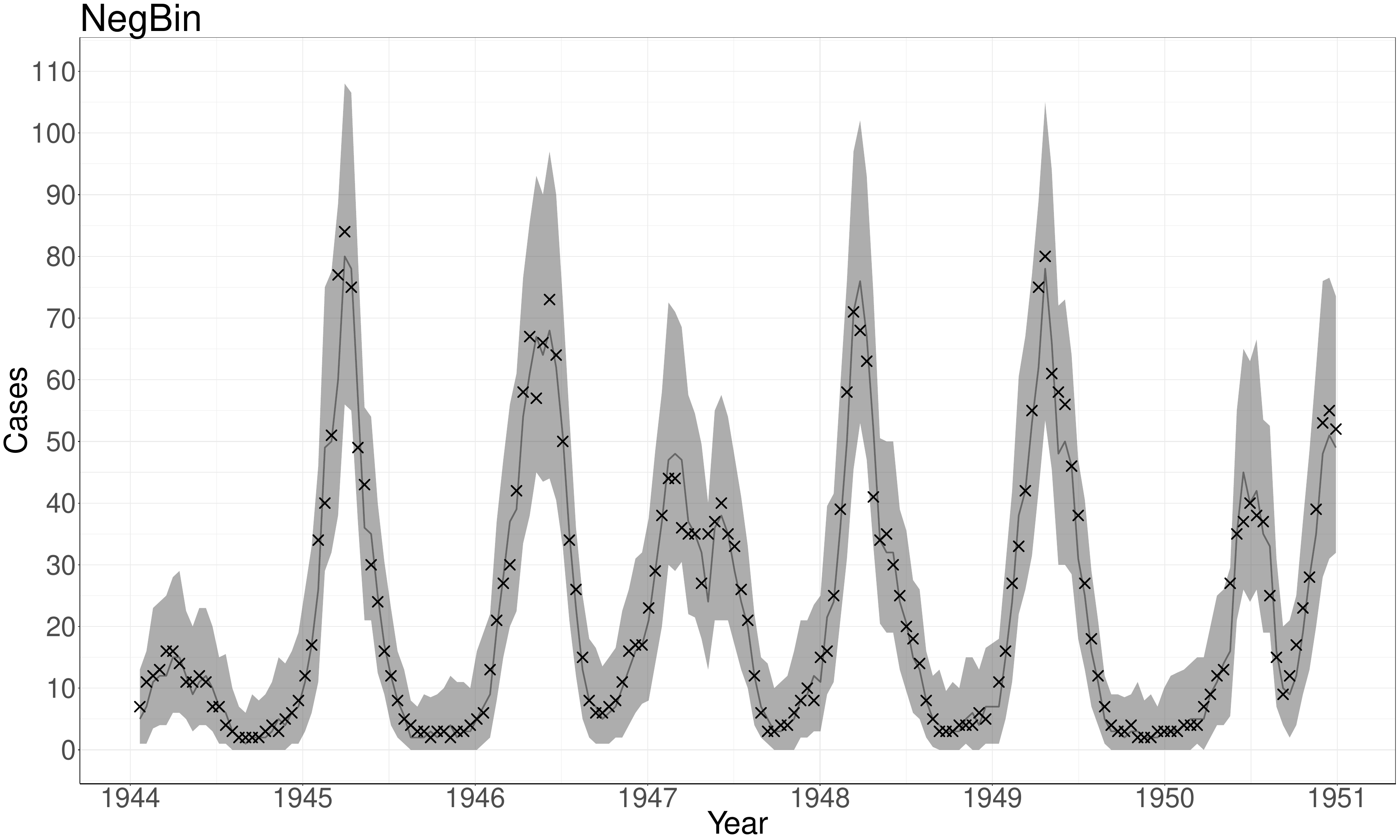}
  \includegraphics[width=1\linewidth]{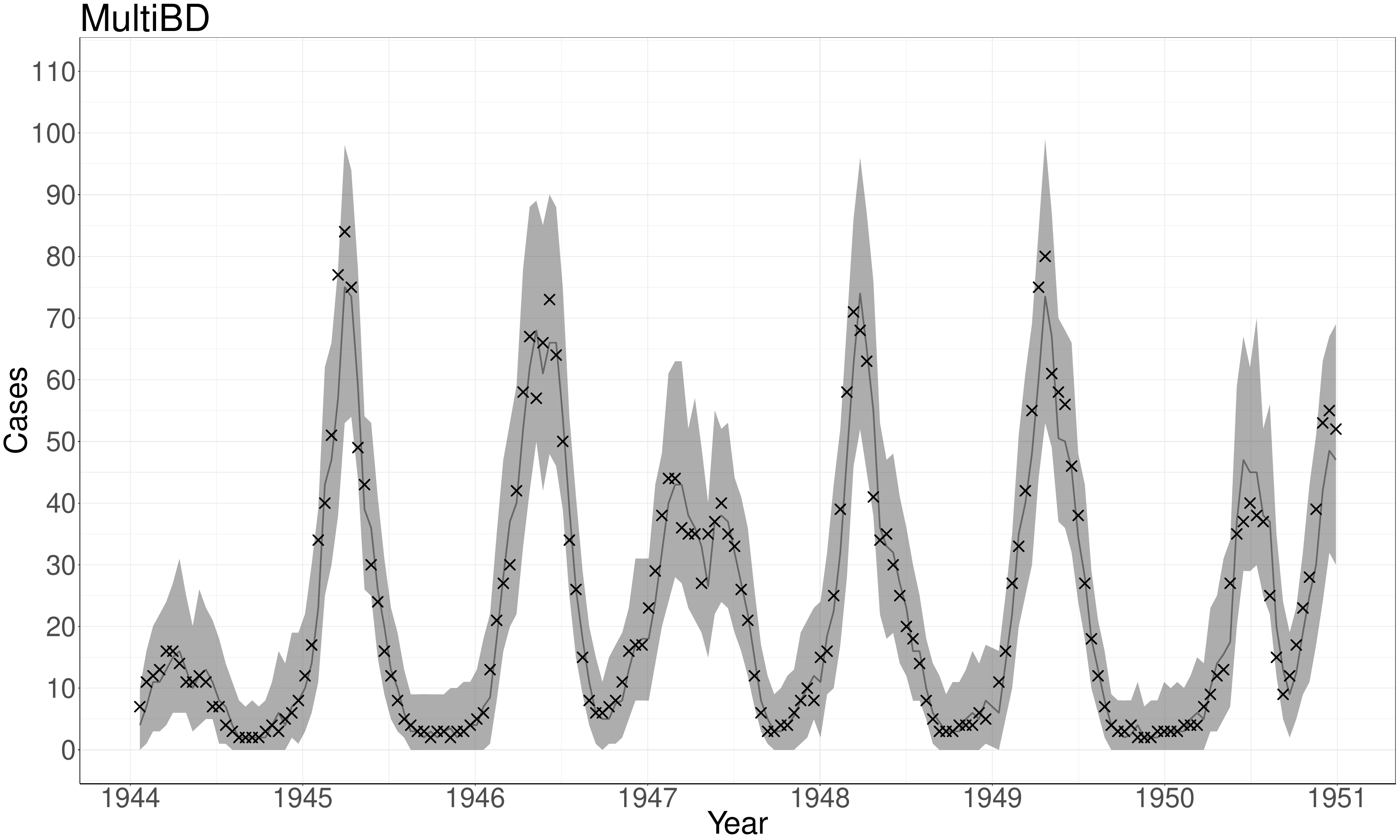}
  \caption{Posterior predictive checks comparing BayesTSIR vs BayesSIR methods}
  \label{fig:PostPredDist}
  \end{figure}

\end{document}